\documentclass[prd,twocolumn]{revtex4}
\usepackage{graphicx, epsfig}
\usepackage{color}
\usepackage{mathrsfs}
\usepackage{amsmath}

\newcommand{\be}{\begin{equation}}
\newcommand{\ee}{\end{equation}}
\newcommand{\bea}{\begin{eqnarray}}
\newcommand{\eea}{\end{eqnarray}}

\newcommand{\gapp}{\mathrel{\raise.3ex\hbox{$>$}\mkern-14mu
              \lower0.6ex\hbox{$\sim$}}}
\newcommand{\lapp}{\mathrel{\raise.3ex\hbox{$<$}\mkern-14mu
              \lower0.6ex\hbox{$\sim$}}}

\begin{document}
\title{Neutralino dark matter stars can not exist}
\author{De-Chang Dai and Dejan Stojkovic}
\affiliation{HEPCOS, Department of Physics,
SUNY at Buffalo, Buffalo, NY 14260-1500}


\begin{abstract}

\widetext
Motivated by the recent ``Cosmos Project" observation of dark-matter concentrations with no ordinary matter in the same place, we study the question of the existence of compact objects made of pure dark matter. We assume that the dark matter is neutralino, and compare its elastic and annihilation cross sections. We find that the two cross sections are of the same order of magnitude. This result has a straightforward and important consequence that neutralinos comprising a compact object can not achieve thermal equilibrium. To substantiate our arguments, by solving Oppenheimer-Volkoff equation we constructed a model of the star made of pure neutralinos. We explicitly showed that the condition for the thermal equilibrium supported by the Fermi pressure is never fulfilled inside the star. This neutralino state can not be described by the Fermi-Dirac distribution. Thus, a stable neutralino star, which is supported by the Fermi pressure, can not exist. We also estimated that a stable star can not contain more than a few percents of neutralinos, most of the mass must be in the form of the standard model particles.
\end{abstract}


\maketitle

\section{Introduction}

Recent observational data \cite{Riess:2004nr,Peiris:2003ff,Spergel:2006hy} indicate that the universe contains a significant fraction ($22$\%) of dark matter whose origin is still unclear. A possible solution to this problem comes from supersymmetric (SUSY)
models in the form of neutralino \cite{Haber:1984rc}. Neutralino is the lightest supersymmetric partner in SUSY, with the mass of about $100$GeV, and is stable. It interacts with the gravitational and weak interactions only, which indicates that it is "dark". Weak interactions and neutralino mass are sufficient to satisfy the relic density needed to explain the observed portion of the dark matter in the universe.

In a remarkable recent study \cite{Massey:2007wb}, a detailed distribution of dark matter as a function of the redshift in a part of our universe was obtained. These observations indicate that dark matter plays a role of a scaffolding upon which ordinary matter builds structures. However, the observations show that large pockets with only dark matter (and no ordinary matter) also exist.
Then a natural question arises whether compact objects like planets, stars or maybe even larger structures made exclusively of dark matter can exist.

Recent study discussed the existence of stars  powered by annihilation of neutralinos \cite{Spolyar:2007qv}. However, these objects contain mostly baryons and only several percents of dark matter.
Compact objects made of pure dark matter were discussed in \cite{Clavelli:2006kn,Narain:2006kx,Bilic:1998kn,Ren:2006tr}. There it was assumed that the equilibrium of these stars was supported by the Fermi pressure.
It was argued that the size of such a dark matter star is less than $1$m with the mass less than $10^{-4}M_\odot$, if neutralino mass is around $100$GeV \cite{Narain:2006kx}. Since the density inside a neutralino star is very high, neutralinos collide frequently.
These earlier studies used the estimate that identical non-interacting fermions have elastic cross section of about $\pi \lambda^2$, where $\lambda$ is de Broglie thermal wavelength  \cite{Pfenniger:2006rd}. With this assumption, non-relativistic neutralinos have the elastic cross section which is much larger than the weak
interaction cross section. If this claim is indeed correct, than a thermal neutralino star could exist, since neutralino annihilation can be neglected in this case. However, it is unlikely that the neutralino elastic cross section is as large as $\pi \lambda^2$. For example, the large elastic cross section would directly contradict the recent "bullet cluster" observations \cite{Clowe:2006eq}.  It is therefore very important to explicitly check this fact.

In order to achieve a thermal distribution, the lifetime of a particle has to be much longer than the characteristic interaction time. Stable neutralinos could in principle achieve thermal state in cosmological context. However, at high densities annihilation can not be neglected. A necessary condition for the thermal equilibrium among neutralinos (at densities smaller than $(100$GeV$)^4$) is that the neutralino-neutralino elastic cross section is much larger than the neutralino-neutralino annihilation cross section.

\section{Neutralino elastic and annihilation cross sections}

In this section, we will explicitly calculate and compare the neutralino-neutralino elastic and annihilation cross sections.
We will show that they are comparable, i.e. of the same order of magnitude. This implies that neutralinos can not be in a thermal equilibrium in a compact object. After several encounters they will annihilate and be converted into the standard model particles. The thermal state of neutralions can not be reached. Thus, the neutralinos can not be described by the the Fermi-Dirac distribution. This then indicates that compact objects made only of neutralinos and supported by the Fermi pressure can not exist.

In SUSY, the neutralinos $\chi^0_i$ are the mixed states of the neutral higgsinos, $\Psi^{1}_{H_1}$ and $\Psi^2_{H_2}$, and
the neutral gauginos, binos ($\lambda_B$) and winos ($\lambda^3_A$) \cite{Rosiek:1995kg}. The mixing and mass matrices are given by

\[ Z^T_N\left( \begin{array}{cccc}
M_1 & 0 & \frac{-ev_1}{2c_W} &\frac{ev_2}{2c_W}\\
0 & M_2 & \frac{ev_1}{2s_W} & \frac{-ev_2}{2s_W}\\
\frac{-ev_1}{2c_W} & \frac{ev_1}{2s_W} & 0 & -\mu\\
\frac{ev_2}{2c_W} & \frac{-ev_2}{2s_W} &  -\mu & 0
\end{array} \right) Z_N \]

\begin{equation}
=\left( \begin{array}{cccc}
M_{\chi_1^0} & 0 & 0 &0\\
0 & M_{\chi_2^0} & 0&0\\
0& 0 & M_{\chi_3^0} & 0\\
0 & 0 &  0 & M_{\chi_4^0}
\end{array} \right)
\end{equation}
where $Z_N$ is the matrix that diagonalizes the mass matrix. Parameters $M_1$, $M_2$ and $\mu$ are mass parameters in the potential, $v_1$ and $v_2$ are the vacuum expectation values of the two higgsions, $s_W$ and $c_W$ are sinus and cosinus of the weak angle, while $e$  is the electron charge. The matrix $Z_N$ also defines the mixing between the states
\begin{eqnarray}
\lambda_B&=&iZ^{1i}_{N}\kappa^{0}_i\nonumber\\
\lambda_A^3&=&iZ^{2i}_{N}\kappa^{0}_i\nonumber\\
\Psi_{H_1}^1&=&iZ^{3i}_{N}\kappa^{0}_i\nonumber\\
\Psi_{H_2}^2&=&iZ^{4i}_{N}\kappa^{0}_i
\end{eqnarray}
\begin{equation}
\chi^0_i=\left( \begin{array}{c}
\kappa_i^0\\
\bar{\kappa}^0_i
\end{array}\right)
\end{equation}
Here, $M_{\chi_1^0}<M_{\chi_2^0}<M_{\chi_3^0}<M_{\chi_4^0}$.

Neutralino $\chi^0_1$ is presumably the lightest supersymmetric particle, and can not decay into other supersymmetric particles. Because of the R-parity conservation, it can not decay into the standard model particles either. Therefore, neutralino is stable. The only way to destroy neutralinos is annihilation.

The cross section for the two-particle final state is
\begin{equation}\label{cs}
\frac{d\sigma}{d\Omega}=\frac{|P_1|}{16\pi^2 E_A E_B |v_A-v_B| E_{cm}}|M(P_A,P_B\rightarrow P_1,P2)|^2
\end{equation}
In our case, both particle A and particle B are neutralinos. In an elastic collision, particles $1$ and $2$ are neutralinos.
In an annihilation process, particles $1$ and $2$ are the standard model particles. $P_1$ is a momentum of the particle $1$. $v_A$ and $v_B$ are velocities of particles A and B respectively. $E_A$ and $E_B$ are energies of particles $A$ and $B$ respectively. $M$ is the scattering amplitude. If the two incoming neutralinos have non-relativistic velocities, say $v\sim 100$km/s, then the momentum $P_1$ of the light standard model particle (except maybe for the top quark) in an annihilation process is about $v/c\sim 1000$ times larger than the momentum $P_1$ of the neutralino in an elastic collision. Thus, only for highly relativistic incoming neutralinos elastic collisions will have $P_1$ of the same order of magnitude as an annihilation process. Since for various different channels all of the parameters are fixed, except for the amplitude, in what follows we compare the amplitudes only.

\begin{figure}[t]
   \centering
\includegraphics[width=3.2in]{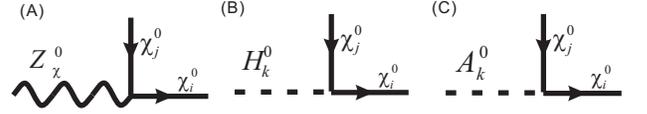}
\caption{The possible vertices of neutralino-neutralino interaction.}
    \label{vertex-neutralino}
\end{figure}

\begin{figure}[t]
   \centering
\includegraphics[width=3.2in]{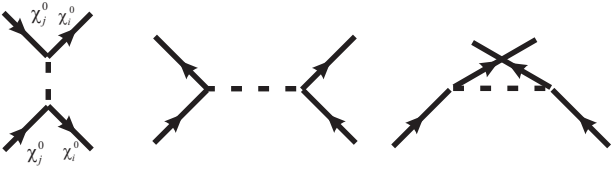}
\caption{The neutralino-neutralino elastic collisions: The dashed line can be $Z^0_\mu$, $H^0_k$, and $A^0_k$.
From the top down are s-, t- and u-channel respectively.}
    \label{elastic}
\end{figure}

We now look for the possible neutralino-neutralino elastic interactions. The possible neutralino-neutralino-something
vertices can be found in \cite{Rosiek:1995kg} and are shown in Fig.~\ref{vertex-neutralino}.
The values of the vertices  are
\begin{eqnarray}
(&A&) :\frac{ie}{2s_W c_W}\gamma^{\mu}\left[(Z^{4i*}_N Z^{4j}_N-Z^{3i*}_N Z^{3j}_N)P_L  \right.\nonumber \\
   &-& \left. (Z^{4i}_N Z^{4j*}_N-Z^{3i}_N Z^{3j*}_N)P_R\right]\\
(&B&):\frac{ie}{2s_W c_W}\{\left[(Z^{1k}_R Z^{3j}_{N}-Z^{2k}_R Z^{4j}_{N})(Z^{1i}_N s_W-Z^{2i}_N c_W) \right.  \nonumber \\
&+& \left. (Z^{1k}_R Z^{3i}_{N}-Z^{2k}_R Z^{4i}_{N})(Z^{1j}_N s_W-Z^{2j}_N c_W)\right]P_L\nonumber \\
&+&\left[(Z^{1k}_R Z^{3i*}_{N}-Z^{2k}_R Z^{4i*}_{N})(Z^{1j*}_N s_W-Z^{2j*}_N c_W) \right. \nonumber \\
&+& \left.  (Z^{1k}_R Z^{3j*}_{N}-Z^{2k}_R Z^{4j*}_{N})(Z^{1i*}_N s_W-Z^{2i*}_N c_W)\right]P_R\}\\
(&C&):\frac{e}{2s_W c_W}\{\left[(Z^{1k}_H Z^{3j}_{N}-Z^{2k}_H Z^{4j}_{N})(Z^{1i}_N s_W-Z^{2i}_N c_W) \right.  \nonumber \\
&+& \left. (Z^{1k}_H Z^{3i}_{N}-Z^{2k}_H Z^{4i}_{N})(Z^{1j}_N S_W-Z^{2j}_N c_W)\right]P_L\nonumber \\
&+&\left[(Z^{1k}_H Z^{3i*}_{N}-Z^{2k}_H Z^{4i*}_{N})(Z^{1j*}_N S_W-Z^{2j*}_N c_W)\right. \nonumber \\
&+& \left.  (Z^{1k}_H Z^{3j*}_{N}-Z^{2k}_H Z^{4j*}_{N})(Z^{1i*}_N S_W-Z^{2i*}_N c_W)\right]P_R\}
\end{eqnarray}
Here, $m_Z = \frac{e}{2s_W c_W} (v_1^2+v_2^2)^{0.5}$ and $m_W=\frac{e}{2s_W}(v_1^2+v^2_2)^{0.5}$. The masses of the Z and W bosons are $m_Z$ and $m_W$ respectively.
From these vertices, the possible Feynman diagrams for elastic neutralino-neutralino interactions are shown in Fig.~\ref{elastic}.
Consider now non-relativistic neutralino-neutralino elastic collision. The amplitude in the s-channel is approximately
\begin{eqnarray}
\label{es-z}
M_{el \, Z^0_\mu}^{(s)} &\sim &  (\frac{e}{s_W c_W})^2\bar{\chi^0_1}\gamma^\mu\chi^0_1  \frac{1}{s-m_Z^2}\bar{\chi^0_1} \gamma_\mu\chi^0_1\\
\label{es-h}
M_{el \, H^0_k, A^0_k}^{(s)} &\sim &  (\frac{e}{s_W c_W})^2\bar{\chi^0_1}\chi^0_1\frac{1}{s-m_H^2}\bar{\chi^0_1}\chi^0_1
\end{eqnarray}
To simplify these equations, omit writing the projectors $P_L$ and $P_R$, since they will not affect the order or magnitude estimate. Here, $s\sim 2M_{\chi_1^0}^2$, while $m$ is the mass of $H^0_k$ and $A^0_k$.
$M_{el \, H^0_k, A^0_k}^{(s)}$ is the amplitude if $H^0_k$ or $A^0_k$ are the mediators, while $M_{el \, Z^0_\mu}^{(s)}$ is the amplitude if $Z^0_\mu$ is the mediator. The t- and u-channel elastic amplitudes are
\begin{eqnarray} \label{el-tu}
M_{el \, H^0_k, A^0_k}^{(t,u)} &\sim & (\frac{e}{s_W c_W})^2\bar{\chi^0_1}\chi^0_1\frac{1}{v^{t,u}-m_H^2}\bar{\chi^0_1}\chi^0_1\\
M_{el \, Z^0_\mu}^{(t,u)} &\sim &  (\frac{e}{s_W c_W})^2\bar{\chi^0_1}\gamma^\mu\chi^0_1\frac{1}{v^{t,u}-m_Z^2}\bar{\chi^0_1}\gamma_\mu\chi^0_1
\end{eqnarray}
$v^{t,u}$ can be $t$ or $u$. In this case $v^{t,u}\sim 0$. If masses of $H^0_k$, $A^0_k$ and $Z^0_\mu$ are comparable to $M_{\chi_1^0}$ (and the energies are not tuned to be in the resonant channel), then all the channels are of the same order of magnitude.

\begin{figure}[t]
   \centering
\includegraphics[width=0.8in]{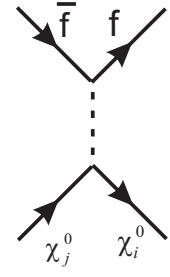}
\caption{The neutralino-neutralino s-channel annihilation: The dashed line represents $Z^0_\mu$, $H^0_k$, and $A^0_k$.}
    \label{annihilation-s}
\end{figure}

\begin{figure}[t]
   \centering
\includegraphics[width=3.2in]{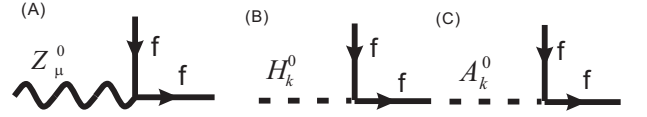}
\caption{The s-channel annihilation vertices.}
    \label{annihilation-vertex}
\end{figure}

We now compare the s-channel elastic scattering amplitude with the s-channel annihilation amplitude (shown in Fig.~\ref{annihilation-s}). For the simple order of magnitude estimate we ignored the production of Z and W bosons. We note that this a conservative approximation since these channels will only enhance conversion of neutralinos into standard model particles. In order to find the annihilation amplitude we need to calculate the three vertices in Fig.~\ref{annihilation-vertex}. The values are
\begin{eqnarray}
(&A&):D\frac{ie}{2s_W c_W}\gamma^\mu (P_L-E s^2_W)\\
(&B&):\frac{i}{\sqrt{2}}Y_f Z^{jk}_R\\
(&C&):\frac{1}{\sqrt{2}}Y_f Z^{jk}_H \gamma_5
\end{eqnarray}
 Parameters $D$, $E$, $Y_f$ and $j$ have different values for different fermions. For $u$-quark we have  $D=-1$, $E=\frac{4}{3}$, $Y_f=\sqrt{2}m_f/v_2$ and $j=2$.
For $d$-quark we have $D=1$, $E=\frac{2}{3}$, $Y_f=\sqrt{2}m_f/v_1$ and $j=1$.
For $e$ we have $D=1$, $E=2$, $Y_f=\sqrt{2}m_f/v_1$ and $j=1$.
For $\nu$ we have $D=-1$, $E=0$, $Y_f=0$ and $j=2$.
The amplitudes are
\begin{eqnarray}
\label{as-z}
M_{ann \, Z^0_\mu}^{(s)}&\sim&(\frac{e}{s_W c_W})^2 \bar{f}\gamma_\mu f \frac{1}{s-m_Z^2}\bar{\chi^0_1}\gamma_\mu\chi^0_1\\
\label{as-h}
M_{ann \, H^0_k, A^0_k }^{(s)}&\sim&(\frac{e}{s_W c_W}) \bar{f^c} f Y_f\frac{1}{s-m_H^2}\bar{\chi^0_1}  \chi^0_1
\end{eqnarray}
where $f^c$ is antiparticle of $f$. Again we omit writing $P_L$ and $P_R$. Eq.~(\ref{as-z}) gives the amplitude if the mediator is $Z^0_\mu$, while
Eq.~(\ref{as-h}) gives  the amplitude if the mediators are $H^0_k$ and $A^0_k$. The amplitude in Eq.~(\ref{as-z}) is of the same order of magnitude as the amplitude for the elastic scattering in Eq.~(\ref{es-z}). The amplitude in
Eq.~(\ref{as-h}) is roughly $(m_f/M_Z)$ times the amplitude for the elastic scattering in Eq.~(\ref{es-h}). If $m_f$ is of the same order of magnitude as $M_Z$, we can conclude that the elastic and annihilation scattering amplitude are also of the same order of magnitude. The only case when this may not happen is when the neutralino is lighter than the top quark and its initial energy is very low. In that case top quark can not be in the final state for kinematic reasons. However, as mentioned earlier, the  cross section also includes the momentum of the emitted particle. Since the next particle in the mass hierarchy is b-quark, which is much lighter than neutralino, the final state particle will have a large momentum. Then the annihilation cross section will be enhanced by a large factor as explained after Eq.~(\ref{cs}). Also, there are more branches in the annihilation process than in an elastic collision since the fermion $f$ in Fig.~\ref{annihilation-s} stands for several species of particles. This is more than enough to overcome the mass suppression factor if the neutralino is lighter than the top quark. Finally, for our argument to work, it is enough that there exists only one annihilation channel which is comparable to the elastic cross section. As we will show, t- and u- annihilation channels have no mass suppression factors regardless of the sign of the  neutralino-top quark mass difference.

\begin{figure}[t]
   \centering
\includegraphics[width=3.2in]{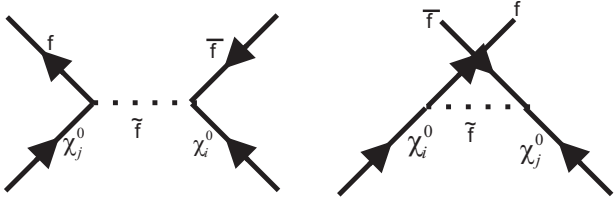}
\caption{The neutralino-neutralino t- and u-channel annihilation: The dashed line represents a sfermion.}
    \label{annihilation-tu}
\end{figure}

\begin{figure}[t]
   \centering
\includegraphics[width=1.2in]{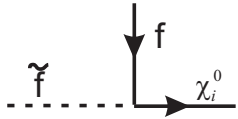}
\caption{The vertices of fermion-neutralino interaction.}
    \label{vertex-tu}
\end{figure}

We now compare the t- and u-channels elastic scattering amplitude with the corresponding annihilation amplitude (shown in Fig.~\ref{annihilation-tu}). In this case, the mediators are sfermions. The vertices needed for these diagrams are shown in Fig.~\ref{vertex-tu}. The values are
\begin{equation}
i(\frac{e}{s_W c_W}A_fP_L+\frac{e}{c_W}B_fP_R)
\end{equation}
where $A_f$ and $B_f$ are constants, which are different for different species \cite{Rosiek:1995kg}. The amplitude is
\begin{equation}
M_{ann}^{(t,u)}\sim (\frac{e}{s_W c_W})^2 \bar{f^c}\chi^0_1\frac{1}{v^{t,u}-m_{\tilde{f}}^2}\bar{f}\chi^0_1
\end{equation}
Again we omitted writing $P_L$ and $P_R$.
Here, $v^{t,u}$ is the square of the momentum transfer, while $m_{\tilde{f}}$ is the mass of sfermion $\tilde{f}$. If $m_{\tilde{f}}$ is not much different from the mass of $Z^0_\mu$, $H^0_k$, and $A^0_k$, then the t- and u-channel annihilation amplitudes are of the same order as the t- and u-channel elastic amplitude in (\ref{el-tu}).

This analysis indicates that the neutralino annihilation amplitude is at least of the same order as the neutralino elastic amplitude. The annihilation cross section can be much larger than elastic cross section, when one includes the momentum difference between the light standard model quarks and heavier neutralinos in the final state, as explained after Eq.~(\ref{cs}).


\section{Model of the star made of pure neutralinos}

In this section we quantify our arguments about the non-existence of the star made of pure neutralinos by treating the Pauli exclusion force as a real interaction between the fermions (different than the weak interaction) .

\begin{figure}[t]
   \centering
\includegraphics[width=3.0in]{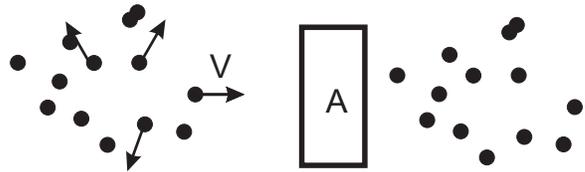}
\caption{The black dots are particles moving with velocity $\vec{v}(E)$. A small wall of area $A$ is embedded inside the particle distribution.}
    \label{collision}
\end{figure}

First we will show that the mean free path for the interaction caused by the Pauli exclusion force must be shorter than the distance between fermions if the Fermi pressure is to provide stability.
To show this, we consider a general setup like in Fig.~\ref{collision}. A wall with an area $A$ is embedded in a group of particles (either fermions or bosons). The number density of particles with energy $E$ is $n(E)$. The particles move with velocity $\vec{v}(E)$. The total particle number density, $N$, and energy density, $\rho$, are
\begin{eqnarray}
N=\int n(E)dE\\
\rho = \int n(E)EdE
\end{eqnarray}

The pressure can be calculated from the momentum transfer of particles that hit the area $A$.
In a three-dimensional space, about $1/6$ of the particles move toward one side of the wall and bounce back. The momentum transfer, ${\cal P}$, in the time interval $\Delta t$ is

\begin{equation}
\Delta {\cal P} (E)= \frac{n(E) A Ev^2 \Delta t }{3}
\label{momentum-change}
\end{equation}

The pressure is
\begin{equation}
P=\int \frac{\Delta {\cal P}(E)}{A \Delta t}dE=\int \frac{nEv^2}{3}
\end{equation}

This is a common result in statistical mechanics. Note that the crucial assumption in Eq.~(\ref{momentum-change}) is that every particle that arrives at the wall $A$ is bounced back. In other words, the efficiency of the interaction between the wall and particles is $100\%$. If the wall bounces back only $\gamma(E)$ of the incoming particles (i.e. $1-\gamma(E)$ of the particles pass through the wall) Eq.~(\ref{momentum-change}) must be rewritten as
\begin{equation}
\Delta {\cal P} (E)= \gamma(E)\frac{n(E) A Ev^2 \Delta t }{3}
\end{equation}
In this case, the pressure must be rewritten as

\begin{equation}
P=\int \frac{\Delta {\cal P} (E)}{A \Delta t}dE=\int \frac{\gamma(E) nEv^2}{3}
\end{equation}

The factor  $\gamma(E)$ then measures the departure from the thermal pressure ($\gamma(E) =1$ means the pressure is equal to the thermal pressure).

Now consider the wall made of particles themselves. From the number density, the size of a particle is $l=n^{-1/3}$. The crucial factor $\gamma (E)$ can be estimated from the ratio between the interaction cross section between the particles, $\sigma (E)$, and the area $l^2$:

\begin{equation}
\gamma (E)= \frac{\sigma (E)}{l^2}=\frac{l}{\lambda (E)}
\end{equation}
Here, $\lambda= 1/(n\sigma) =l^3/\sigma $
is the mean free path of the particles. If $\lambda < l$, i.e. the mean free path is shorter than the distance between the particles, then we need to set $\gamma(E)=1$. However, if the mean free path is larger than the distance between the particles, $\gamma (E) <1$. In that case, the pressure is smaller than the thermal pressure. In the extreme case when $\lambda \gg l$ the particles practically do not interact with each other, i.e. $\gamma(E) \approx 0$, and the pressure can not be built up at all. These arguments are true both for bosons and fermions. For fermions, the Pauli exclusion force builds up the Fermi pressure between the fermions. However, for this to happen, the mean free path for fermions must be smaller than (or at most equal to) the distance between the fermions. We will use this result in what follows.

We now model the structure of the star made of pure dark matter.
The relativistic mean field approximation \cite{Serot:1984ey} implies that the density, pressure and number density of neutralinos can be written as (for details see \cite{Ren:2006tr})
\begin{eqnarray} \label{prk}
\rho&=&\int_0^{k_F}\frac{\sqrt{k^2+m^{*2}}k^2dk}{\pi^2} \\
&&+\frac{m_H^2H^2}{2}+\frac{m_h^2h^2}{2}+\frac{m_z^2Z^2}{2}\nonumber \\
P&=&\int_0^{k_F}\frac{k^4dk}{3\pi^2\sqrt{k^2+m^{*2}}}\nonumber \\
&&-\frac{m_H^2H^2}{2}-\frac{m_h^2h^2}{2}+\frac{m_z^2Z^2}{2} \nonumber\\
N&=&\frac{k_F^3}{3\pi^2} \nonumber
\end{eqnarray}
Here, $k_F$ is the Fermi momentum.
 Also
\begin{eqnarray}
H&=&\frac{-g_H}{m_H^2 \pi^2}\int^{k_F}_0\frac{m^{*2}}{\sqrt{k^2+m^{*2}}}k^2dk\\
h&=&\frac{-g_h}{m_h^2 \pi^2}\int^{k_F}_0\frac{m^{*2}}{\sqrt{k^2+m^{*2}}}k^2dk \nonumber\\
Z&=&\frac{g_z}{3m_z^2\pi^2}k_F^3 \nonumber \\
m^*&=& m_\chi +g_H H+g_h h  \nonumber
\end{eqnarray}
 The numerical values of the relevant parameters are $m_\chi=1.01679763\times 10^6 MeV$, $m_H=7.5336438\times 10^5 MeV$,
$m_h=1.18760345\times 10^5 MeV$, $m_z=9.1187\times 10^4 MeV$, $g_H=3.131\times 10^{-1}$, $g_h=-2.705\times 10^{-2}$, and $g_z=1.43\times 10^{-3}$.

The neutralino annihilation cross section is $\sigma_a \sim 1\times 10^{-10} MeV^{-2}$. The mean free path for annihilations is $\lambda _a =1/(N\sigma_a)$. As we showed above, the mean free path of the Pauli exclusion force is about the distance between fermions, i.e. $\lambda_p =N^{-1/3}$.
To keep neutralinos in thermal equilibrium, we need
\begin{equation} \label{cond}
\frac{\lambda_p}{\lambda_a}=N^{2/3}\sigma_a <<1
\end{equation}

\begin{figure}[t]
   \centering
\includegraphics[width=3.2in]{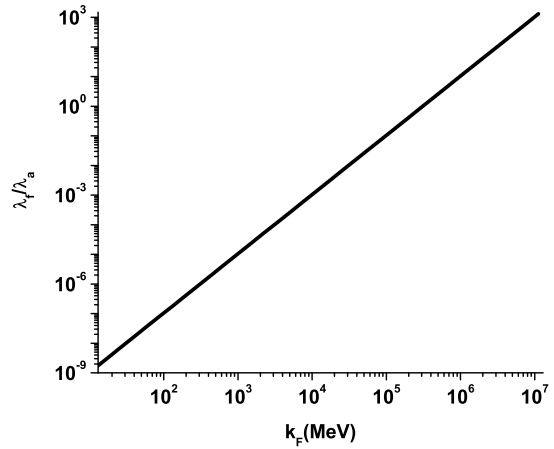}
\caption{The dependence of the ratio between the mean free paths for the Pauli exclusion force and annihilations $\lambda_p/\lambda_a$ on the Fermi momentum $k_F$. The condition (\ref{cond}) breaks down for high $k_F$ (high densities).}
    \label{ratio}
\end{figure}

From Fig.~\ref{ratio}, we see that the condition (\ref{cond}) can not hold at high $k_F$ (i.e. high densities).

We now write the Oppenheimer-Volkoff equation \cite{Oppenheimer:1939ne}
\begin{eqnarray} \label{structure}
\frac{dP}{dr}&=&-\frac{(\rho +P)(M+4\pi Pr^3)}{r^2 (1-2M/r)},\\
\frac{dM}{dr}&=&4\pi \rho r^2 \nonumber .
\end{eqnarray}
Combining Eqs.~(\ref{structure}) with the equation of state given by Eqs.~(\ref{prk}), we can find the structure of the neutralino star. We need to setup the boundary condition for $k_F$, so we choose $k_F=0.5 m_\chi$ at the center of the compact object. We then numerically integrate Eqs.~(\ref{structure}). The radius of the star is the distance from the center at which the pressure drops to zero (in this case the radius of the star is $1.2$cm). This gives us the structure of the star, i.e. energy density, pressure and neutralino number density as functions of the distance from the center of the star.

\begin{figure}[t]
   \centering
\includegraphics[width=3.2in]{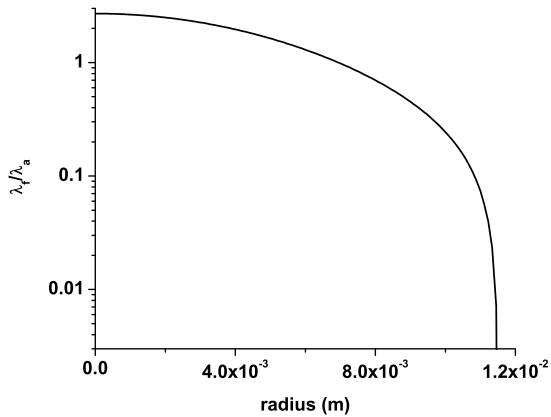}
\caption{The dependence of the ratio between the mean free paths for the Pauli exclusion force and annihilations $\lambda_p/\lambda_a$ on the distance from the center of the compact object made of pure dark matter. Throughout the star the condition (\ref{cond}) is violated. The ratio $\lambda_p/\lambda_a$ drops sharply only near
the surface of the star.}
    \label{ratio-r}
\end{figure}

Fig.~\ref{ratio-r} shows the ratio of  $\lambda_p/\lambda_a$ vs. radius. Clearly, the ratio is about $1$ throughout the star. The ratio drops sharply only at the surface of the star. This indicates that the Pauli exclusion force is not sufficient to overcome annihilation. The neutralinos decay into the standard model particles before the star becomes thermal.

One may roughly estimate the ratio between the standard model particles and neutralinos which is needed to stabilize the star. This ratio will depend on the mean free path for neutralino annihilation and that of the standard model particles. For a condition (\ref{cond}) to hold, the mean free path of the standard model particles can not be greater than several percents of the mean free path for neutralino annihilation. This implies that the neutralino content in the star can not be greater than a few percents.

Finally, we comment on the interplay between gravity and annihilations. Gravity brings neutralinos together, but annihilations put an upper limit to the density of (pure) dark matter clumps. Even without detailed numerical simulations we can perform a simple order of magnitude estimate. Consider for example a neutralino clump that was presumably created when the galaxy was formed, and survived till now. The density of the clump can be calculated from the condition
\begin{equation}
\Delta T<\frac{\lambda_a}{v_\chi}=\frac{1}{\sigma_a N v_\chi}
\end{equation}
where $\Delta T$ is the lifetime of a galaxy --- about $10$ Gyrs, while the velocity of dark matter is assumed to be $100$ km/s. $N$ is the total number density of neutralinos and the neutralino annihilation cross section is $\sigma_a \sim 1\times 10^{-10} MeV^{-2}$. From this condition we find that the number density of the clump must be lower than $N\sim 6.2\times 10^{-35}$ GeV$^3$ $\approx 8.2\times 10^{12}$ m$^{-3}$, which for a $100$GeV neutralino gives the mass density of $10^{-13}$g/cm$^3$. For more general constraints, one would need to perform detailed numerical simulations of the gravitational collapse with annihilations included.

\section{Conclusions}

In conclusions, we studied here whether pure dark mater can make compact objects like stars. The study was motivated by the remarkable recent observations \cite{Massey:2007wb} that there are regions in our universe with pure dark matter distribution without the visible presence of the ordinary matter.
We explicitly calculated and compared the neutralino elastic cross section with the annihilation cross section and found them to be comparable. This means that, in each encounter, neutralinos have about the same chance to annihilate into the standard model particle or to be scattered. If they are scattered, they exchange energy and momentum and after many repeated encounters will reach the thermal equilibrium. However, annihilation will prevent this scenario from happening. Once neutralions get converted into the standard model particles they can not come back. This situation can not be described by the Fermi-Dirac equation of state. This implies that a stable neutralino star supported by the Fermi pressure can not exists.

To substantiate our arguments, by solving Oppenheimer-Volkoff equation we constructed a model of the star made of pure neutralinos. We explicitly showed that the condition for the thermal equilibrium
supported by the Fermi pressure is never fulfilled inside the star.
We also estimated that a stable star can not contain more than a few percents of neutralinos, most of the mass must be in the form of the standard model particles. This may have implications for the question why we are made of the standard model particles if most of the matter in the universe comes in the form of dark matter. For intelligent observers like us to evolve, structures like stars and planets are (perhaps) necessary.

One of the possibilities to have neutralinos in thermal equilibrium is an environment with enormous densities, of the order of $(100$GeV$)^4$. If such an environment is thermalized, then its temperature is of the order of the rest mass of the neutralino.  In this case neutralinos can be produced thermally. This can only happen in early universe or in the central core of a very dense star.
In the case of the central core of a star, this region must be very  tiny, with the radius smaller than $1$cm \cite{dlss}. Again, such an object can not be supported by the Fermi pressure. The annihilation process would be so quick that the star could not exist long enough to be observed.

The other possibility to have neutralinos in thermal equilibrium is an environment with very low densities. In an environment where the neutralino density is so low that neutralinos barely collide with each other, their effective lifetime is much longer than the weak-scale characteristic collision time.
In such conditions, gravitational cooling may bring neutralinos to equilibrium. This is the mechanism behind dark matter halo formation. However, these are rather low density dark matter distributions and can not be called compact objects.
At higher densities gravitational scattering can not compete with weak interactions. Therefore, gravitational cooling may bring neutralinos together but can not play an important role in the subsequent evolution of the compact object.

We assumed here that the dark matter is comprised of neutralinos.
However, any particle with the mass around the weak energy scale and weak scale interaction cross section (such is neutralino) automatically has a relic density which satisfies observational constraints. Thus, we expect that the same conclusions will hold for more general models. Therefore, despite the fact that regions in our universe with the pure dark matter distribution were observed, compact objects like planets, stars and larger systems like galaxies made of dark matter are unlike to exist.

We also note that we used the dark matter models in their most standard form. In some exotic models, dark matter particles with an extra self-interaction \cite{Spergel:1999mh,Burkert:2000di,Ackerman:2008gi,Feng:2008ya,Hooper:2008im,Kim:2008pp,Huh:2007zw,Dai:2009hx} were considered to solve the problems with subhaloes \cite{Moore:1999wf, Ghigna:1999sn}, cuspy cores \cite{Navarro:1995iw,Moore:1994yx,Flores:1994gz,de Blok:1997ut} and recent apparent PAMELA \cite{Adriani:2008zr}, ATIC \cite{:2008zzr} and FERMI \cite{Abdo:2009zk} observations. In the context of these models, this extra self-interaction has to be properly included.


\begin{acknowledgments} The authors are grateful to D. Wackeroth, A. Scharf, Y.C. Lee, G. Starkman and H. Mathur for very useful conversations. DS acknowledges the financial support from NSF.
\end{acknowledgments}

\end{document}